\documentclass[manuscript, screen,nonacm]{acmart}

\AtBeginDocument{%
  \providecommand\BibTeX{{%
    \normalfont B\kern-0.5em{\scshape i\kern-0.25em b}\kern-0.8em\TeX}}}

\usepackage{enumerate}
\usepackage{hyperref}

\setcopyright{none}
\renewcommand\footnotetextcopyrightpermission[1]{} 
\begin{document}

\title{Leveraging Collaboration for Multifaceted Design and Product Teams: A Financial Perspective}
\author{Esha Shandilya}
\authornote{Both authors contributed equally to this research.}
\email{eshandilya@chathamfinancial.com}
\author{Jacalyn DeFeo}
\authornotemark[1]
\email{jdefeo@chathamfinancial.com}
\affiliation{%
  \institution{Chatham Financial }
  \city{Kennett Square}
  \state{PA}
  \country{USA}
}


\begin{abstract}
  Collaboration is a key driving force for a team’s success. In this case study, we discuss the collaboration practices of the Design Team and a subset of product teams at Chatham Financial—a financial risk management advisory and technology firm. The Design Team’s collaboration workflow has four key elements surrounding the structure, cross-team communication, onboarding, and feedback that occurs both in team and cross-team collaborative partnerships. Each of the key elements lead to a unique set of challenges and opportunities for the Design Team. We analyze the current state of each element, their value proposition, challenges, and initiatives undertaken to make the collaboration practice more robust.

\end{abstract}




\maketitle
\section{Introduction}
\label{sec:introduction}
Chatham is an employee-owned, purpose-driven organization focused on positively impacting the markets, our clients, our employees, our shareholders and the world and our communities~\cite{chatham}. Providing an impactful positive experience to our clients depends on gaining client awareness and empathy, which is essential for the Design Team at Chatham to be able to design intuitive user experience solutions~\cite{lin2007empathic,wright2008empathy}. Being a Design Team in a financial-risk-tech space is a unique position to be in because, for us, collaboration and communication with business and product teams are critical tools to gain business and product knowledge and build client empathy. Successful collaborations happen when diverse teams like Design and business and product are aware of each other's existences, avenues to collaborate, and ways to leverage each other's expertise to the fullest potential~\cite{young2007knowing}. Therefore, the Design Team at Chatham focused on understanding how we currently collaborate as a Design Team with product teams across the organization, identifying gaps and opportunities within our collaborative approach, processes, and strategies to improve. In this case study, we discuss how, through our internal research study, we uncovered four elements of our collaboration workflow: 1) Centralized Squad Structure, 2) User Experience Squad Connects Design \& Product, 3) Project Request Notification \& The Onboarding Situation, and 4) Feedback \& Mentorship, and identified the promoters and opportunities to support conducive collaboration. Our current collaborative framework with its four elements is unique to Chatham and marks us as a distinct Design Team.

\section{Design \& Collaboration at Chatham Financial }
\label{sec:background}
\begin{figure}[]
      \centering
      \includegraphics[scale=0.22]{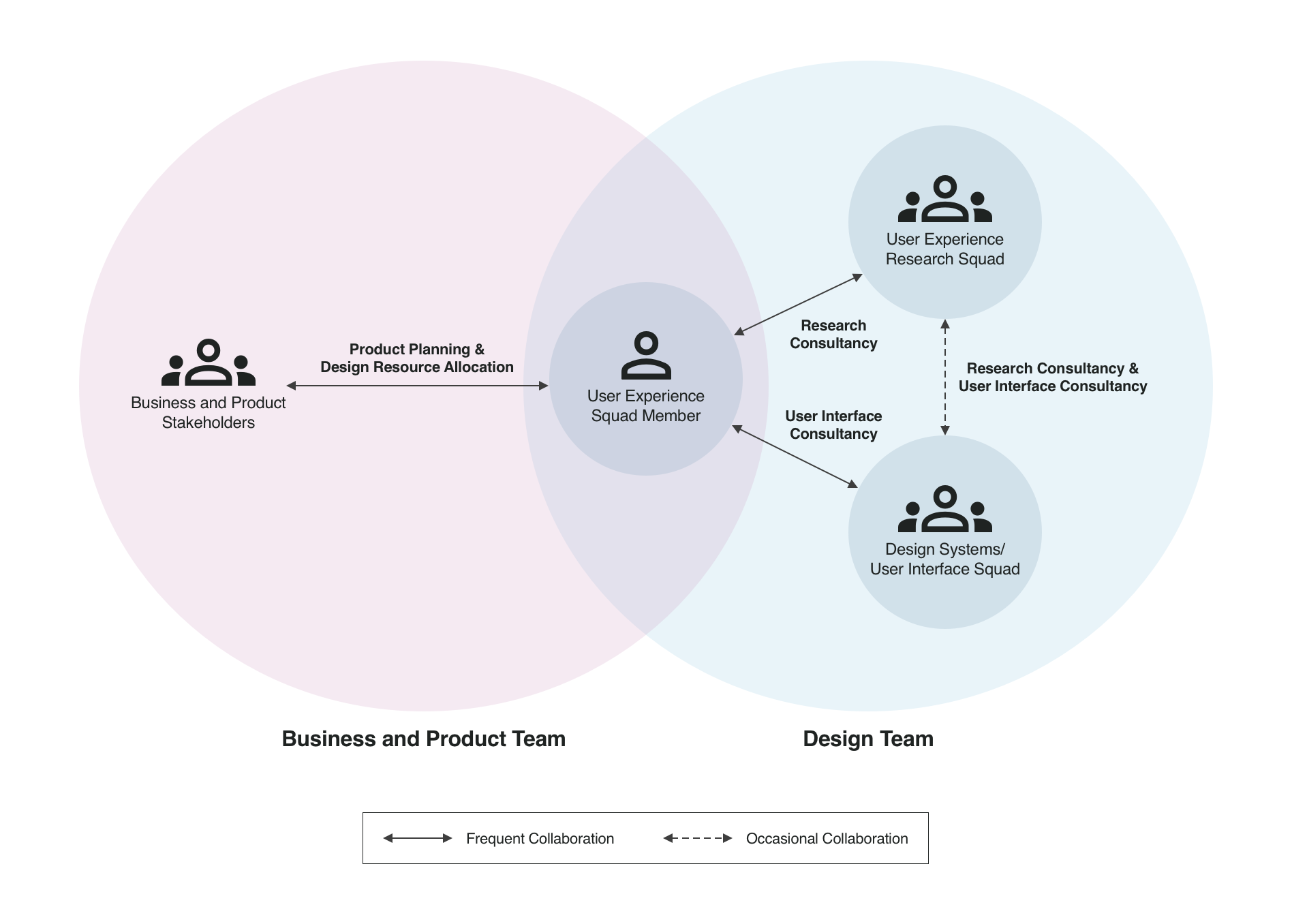}
      \caption{This snapshot provides a quick overview of how squads within and outside of the Design Team collaborate together. The User Experience Squad Member acts as an intermediary between the Business and Product Team and the Design Team.}
      \label{fig:structure}
\end{figure}

The Chatham Design Team is comprised of four main squads. The Design System/User Interface (DS/UI) squad helps enhance product design aesthetics, consistency, and accessibility, while managing and documenting components that are utilized across products for multiple business teams. The User Experience Research (UXR) squad is responsible for implementing numerous research methods to gain empathy for the user’s pain points and define core deliverables for a product. The User Experience (UX) squad works hand-in-hand with business and product stakeholders to help determine product requirements, design product interactions, and act as an intermediary between the Design Team and business and product team, as shown in Fig. \ref{fig:structure}. While these teams frequently collaborate to pursue a common product goal, the teams operate in varying ways. The fourth squad, the Design Operations squad, was newly introduced when this study was conducted, and therefore, we only focused on the collaboration between the three main squads—DS/UI, UXR, and UX.\looseness=-1

Each UX squad member is deeply embedded in a business and product team—after receiving a project notification from the business and product stakeholder, they work constantly with them to not only understand the requirements but are also responsible of allocating work to the centralized squads. Upon being notified of a project from the UX squad member, the DS/UI and UXR squads act as consultants throughout the project duration. The UXR squad helps each UX member gain insight into the problem they are trying to solve by helping them gain empathy and defining user groups. The DS/UI squad helps the UX member determine the visual aspects of an interface by promoting consistency from different interfaces and components they have maintained. From our findings, which we discuss later in this case study, we found that not only is there a stark contrast between the embedded nature of each UX squad member and the centralized nature of the DS/UI and UXR squads, but also between how individual Design Team members collaborate. 

Each member of the Design Team carries a unique skill-set and perspective on collaboration, but all have a desire to work together and accomplish a common goal. Even so, creating a conducive collaborative structure would enable efficient collaboration standards. Currently when squads collaborate, they spend too much time trying to work together efficiently due to a lack of clear expectations or processes for how to do so efficiently. There is also reliance to gather and extract critical data points from each UX squad member. This opens up an opportunity to make improvements that better equip each UX squad member, as well as enhance participation in formal cross squad collaboration. Implementing this conducive structure would allow the Design and business and product teams to leverage the full potential of each squad, while still providing support for unique styles of collaboration. 

\section{Our Research Approach}
\label{sec:method}
The Design Team at Chatham is visionary when it comes to leveraging research to make the internal collaboration processes efficient. Moreover, we value how research helps the team reflect while also bringing awareness to how we collaborate and opportunities to improve. Surveying with the intent of conducting interviews was a step towards reflecting and observing what the current state of the collaboration workflow among different squads and teams looked like. Members of the Design Team were surveyed to understand their perception and to gather brief reflections on collaboration. Candidates were then identified for the interviews to shed light on the pain points collected from the survey. An inductive qualitative analysis was performed on the gathered survey and interview data. 

The survey focused on gaining a deeper understanding of how the Design Team perceived the term “collaboration”. Team members ranged from less than one year of experience to over ten, and included the DS/UI, UXR, UX and Design Operations squad members. Questions focused on 1) how they defined collaboration, 2) their experience collaborating, and 3) examples of collaboration that gave insight into their perspective of collaboration based on their different disciplines. Surveying the entire team allowed us to get a general understanding of the current state of collaboration, while also bringing to light areas that needed further clarification through interviews. This also informed the decision on the candidates we interviewed, as we discovered through our survey that the Design Team had three levels of collaboration frequency—frequent, semi-frequent, and seldom. While this survey was solely conducted on the Design Team members, we could have extended the survey out to the business and product teams to get a deeper understanding of their collaborative landscape prior to interviews. 

Team leads for the DS/UI and UXR squad along with three UX squad members and two business and product stakeholders were interviewed to understand their pain points and needs around efficient collaboration. Out of the three UX squad members, each had varying levels of collaboration frequency. Both business and product stakeholders interviewed work closely with each of the UX squad members that were recruited for the interviews. Questions spanned 1) the current process of collaborating with the Design Team, 2) understanding the problem space, and 3) understanding future ideal collaboration. With collaboration being unique to members both within and outside of the Design Team, the business and product stakeholders interviewed may have their own varying level of collaboration frequency compared to other potential stakeholder candidates.  

While we pride ourselves in being dynamic and adaptive to improvements, this data captured serves as a snapshot in time amongst the beginning of initiatives being implemented. 

\section{Findings \& Opportunities: The Four Elements of Our Collaboration Workflow }
\label{sec:results}
In this section, based on our learnings gathered from the internal research study, we discuss how the Design Team collaborates at Chatham to gain empathy for end-users of our niche financial products. After performing an inductive qualitative analysis, we discovered four elements of our collaboration workflow: 1) Centralized DS/UI \& UXR Squad Structure, 2) UX Designers Connect Design \& Product, 3) Project Request Notification \& The Onboarding Situation, and 4) Feedback \& Mentorship. 

\subsection{Centralized DS/UI \& UXR Squad Structure} 

Chatham is committed to its clients’ success by creating and delivering consistent and optimal product experiences. The Design Team comprises independent squads with DS/UI, UXR, and UX expertise to support Chatham's values. The UX squad is embedded into each business and product team; refer to Fig. \ref{fig:structure}, which helps centralized squads like DS/UI and UXR get quick updates and notifications about the business teams’ latest trends and news. Our structural setup allows squads to provide consistency in design solutions, execute new research strategies, and identify avenues for intra-squad learning and collaborations. 

Because of this, centralized squads constantly invest their efforts and skills to deliver countless active projects. Consequently, exposing the centralized squads to various challenges: 
\begin{enumerate}
    \item {\em Limited expertise in a business and product vertical:} Centralized squads, DS/UI and UXR, have limited time to understand the business requirements and background knowledge of a project due to simultaneously consulting on multiple projects. As a result, the centralized squads can be deprived of acquiring domain expertise in a particular business and product vertical. Moreover, other projects get sidelined if a centralized squad dedicates its effort to a complex undertaking. 

\item {\em UXR's impact:} Being a centralized squad, UXR has limited avenues to collaborate and learn about business, product, and market trends, therefore, UXR is often not part of diagnosing a problem. They are notified of a project requiring research efforts with a pre-determined research method by UX squad members, such as designing and conducting usability test sessions, because those methods have been proven to be successful in the past. As a result, the centralized structure reduces the UXR squad's impact that they can bring to the table if they are involved in the product's visions and goals.  

\item {\em Lost collaborative partnerships with stakeholders:} Centralized squads, DS/UI and UXR, lose the opportunity to build collaborative partnerships with the business and product stakeholders, like product managers and subject matter experts (SMEs), as these squads do not disseminate their deliverables and impact directly to them. Understanding how squads like UXR and DS/UI contribute is crucial for business and product stakeholders to comprehend how to leverage the centralized squads' expertise in the future. 
\end{enumerate}

Below are a few initiatives that the Design Team at Chatham is progressively working towards to address the gaps: 

\begin{enumerate}
    \item {\em Sharing, learning, and winning opportunities:} Centralized squads frequently share the design deliverable case studies with various business stakeholders and product teams to broadcast how squads contributed and added value. This exercise has been helpful for stakeholders outside of Design team to understand ways in which a particular squad can help and recognize opportunities to connect with the squad directly for their expertise.

\item {\em Understand stakeholders' perceptions of design and research to introduce design initiatives:} Recognizing a need for a potential squad's efforts is winning a half battle; the other half lies in devising a framework to collaborate effectively. The UXR squad acknowledges this and has started separate research studies with respective business teams to understand how design is perceived and the potential for building effective collaborative partnerships with non-design stakeholders. 

\item{\em Prioritize, plan, and act:} Our centralized squads invest heavily in planning project requests. However, the influx of project requests due to our team structure demands having a priority parameter in place that could inform the amount of time dedicated to a particular project. Moreover, embedded UX designers consciously try bringing together all the respective members of the squad and stakeholders (product owner, developer, and SMEs) during the planning phase. This initiative has allowed squads and stakeholders to represent their thoughts, get pushback for cost and minimal viable product (MVP), project requirements, and discuss the scope. 
 \end{enumerate}

\subsection{UX Squad Connects Design \& Product} 

As mentioned above, the UX squad comprises designers embedded into each business and product team. Besides designing high-impact user experiences for Chatham's products and services, UX squad members also connect the worlds of the Design Team to the respective business and product teams. In addition, UX squad members are the most prominent advocators of the Design Team and their contributions at Chatham. They apprise and persuade the business and product teams when and how to utilize the centralized squads and themselves for a project, and them for a project. UX squad members collaborate most frequently with their product owners/managers to understand business concepts, user groups and SMEs, product requirements, and how it applies to the project space. Once UX squad members are notified of a problem, they try to flesh out the needs of that project and before connecting with the respective centralized squads to share product knowledge and product people dynamics/preferences, which helps the centralized squads to estimate and strategize their work.  

In the current workflow, the UX squad helps the centralized squads gather business and product knowledge. However, the workflow posed various implications, mainly for the business and product stakeholders and the UX squad: 
\begin{enumerate}
    
\item{\em Business and product teams' limited perception of design and its offerings:} Product owners/managers depend on the UX squad’s design knowledge and discretion to onboard or use the skills of the DS/UI and UXR squads at specific product life cycle phases. This workflow deprives business and product owners/managers of exposure to the Design Team's service offerings through firsthand collaborations. Moreover, how, when, and what the centralized squads can help with. This knowledge gap of business and product teams may also restrict future collaborations with them and the Design Team beyond the UX squad.  

\item{\em The UX squad and their herculean collaborative efforts inside and outside the Design Team:} The UX squad is the knowledge hub of information sharing ranging from squads' skills to project and business requirements. If in any situation, due to project deadlines, they skip notifying or gleaning the required information crucial for a squad, it may result in a lot of back and forth, causing wasted time and effort for the entire project. 
\end{enumerate}
To enhance the business and product teams’ knowledge of the Design Team roles, while also helping the UX squad in their collaborative efforts, we are creatively re-introducing the Design Team's squads and their responsibilities (e.g., using analogies to describe the primary job duties of a squad and project success stories) in company-wide meetings. This exercise will help the business and product teams understand how the Design Team is structured and in what ways they can contribute to enhancing the user experiences in product offerings and beyond.

\subsection{Project Request Notification \& The Onboarding Situation }

As discussed above, in most cases, the UX squad notifies the centralized squads of upcoming project requests. In this section, we will discuss the various current ways project request notifications occur, as well as how squads within the Design Team are both notified and onboarded to new and existing projects

\textbf{UXR Squad} 

For the UXR squad, the project request notifications happen in the following ways: 
\begin{enumerate}
\item{\em UX ---> UXR:} The UX squad reaches out to the UXR squad with specific requests like usability tests or interviews, whereas other times, the requests are more about understanding the problem space. The UX squad usually informs the UXR squad members of projects through Slack or quick in-person meetings. 

\item{\em UXR ---> UX and DS/UI:} The UXR squad also uses proactive check-ins with the UX and DS/UI squads to get notified of any upcoming research projects and provide consultations to designers to tackle a specific problem.  

\item{\em Product Owner, DS/UI ---> UXR:} Although rare, in a few situations the UXR squad has an opportunity to get notified directly by a product owner/manager. Similarly, the DS/UI squad may reach out to the UXR squad with specific research requests like designing test sessions for UI testing, etc.   
\end{enumerate}

\textbf{DS/UI Squad}

For the DS/UI squad, the project request notifications happen in the following ways: 
\begin{enumerate}
\item{\em UX ---> DS/UI:} The UX squad contacts the DS/UI squad when they want feedback or ideas around a particular design problem or want them to review the guidelines for a UI component, as the DS/UI team follows a contribution model to maintain the Design System's content. The UX squad also invites the DS/UI squad to collaborate on designing interfaces/concepts together.  

\item{\em UXR ---> DS/UI:} Although rare, in a few situations the UXR squad hands off a project to the DS/UI squad once research insights on a problem are ready, and the UXR squad needs their expertise to visualize the insights through design deliverables like personas. 
\end{enumerate}

The UX squad is the primary point of contact for the UXR and DS/UI squads and are responsible for keeping themselves abreast of the respective business teams’ latest trends and news. This helps the centralized squads strategize and anticipate their efforts and contributions to support the business objectives. However, the current process of request notification and onboarding a squad into an ongoing project exposes them to multiple challenges: 

\begin{enumerate}
\item{\em Unstructured process of squad onboarding to an ongoing project:} Usually, due to a high volume of project request notifications, when a centralized squad service is requested, there is limited to no onboarding at all, and there are no introductory sessions scheduled to provide the background of a project. To elaborate, squads rely on the UX squad’s knowledge to extract the relevant details for the project, from business knowledge to understanding the timing and deliverables of the project. Centralized squads have to figure out what they are doing because there is no set procedure to onboard the squads. 

\item{\em Centralized squads not included during the project's planning phase:} The centralized squads' onboarding does not happen during the planning phase of the project, resulting in fewer avenues for the Design Team squads to collaborate and contribute effectively. 

\item{\em Self-service to set the stage before getting started:} The centralized squads find it challenging to figure out the critical aspects of the project, which are essential for them to understand to get started on the project. When the UX squad or requester is unaware of the critical required data points, the squads then have to spend additional time and effort to gather the knowledge and relevant information required to get started on their own. In this case, the squads spend additional time and effort to gather the knowledge and relevant information required to get started. For example, sometimes the DS/UI squad finds themselves working with high-level vague problem statements, which shifts their focus and effort from designing to refining the problem statements with the requester. 

\item{\em Cognitive overload on the requester, the UX squad:} The UX squad finds transferring knowledge about an ongoing project to the squads challenging. They find it difficult identifying the relevant information that a particular squad would want to have in place prior to their contribution. 
\end{enumerate}

To alleviate the cognitive overload on the UX squad and introduce a structured collaboration framework that enables effective squad collaboration within and outside of the Design Team, our team is working on the following:  

\begin{enumerate}
    \item{\em Squad onboarding template for the Design Team:} We are curating a squad onboarding template for the Design Team, which lists all the information and data points the squads require to get started on a project. After reflecting on our workflows, we have a general list of data points in the onboarding template applicable to all the squads: \looseness=-1
        \begin{enumerate}[(i)]
            \item Setting/conveying clear asks for the squad, for example, what type of service is needed from the squad
            \item Project name and stakeholders  
            \item A clearly defined/documented project brief, mentioning the project name and stakeholders.  
            \item Intended client user group  
            \item Information and evidence of clients' pain points  
            \item List of collaborators/squads/teams involved  
            \item Explanation of any pre-work that has been completed 
            \item If the product/feature already exists or is brand new  
            \item The phase of the product lifecycle the project is in  
            \item Identifying the key milestones and timeline  
            \item Outside of the request from the squad, thoughts on how stakeholders see the project progressing and how the squads' inputs will inform the next steps 
        \end{enumerate}

\item{\em Frequent 1:1 sessions with business and product stakeholders:} Given that our organization is unique in its service offerings, it may be difficult for centralized squads in the Design Team to gain business and product knowledge. To overcome this, we lean on the business and product experts to acquire knowledge; respective squads schedule 1:1 sessions with the experts to identify the business and product knowledge crucial to informing and strategizing the next steps. 

\item{\em Project briefing to inform the research plan for effective collaboration between diverse teams:} Besides the work-in-progress squad onboarding template, the UXR squad also has a project briefing template to inform the research plan for complex discovery projects requiring generative research. The research plan has proven effective for productive collaborations between the UXR squad, UX squad, and business and product teams. The research plan helps diverse teams think through a problem and how to approach it and has made acquiring knowledge easy. 
\end{enumerate}

\subsection{Feedback \& Mentorship}

The Design Team at Chatham acknowledges that collaboration helps break tunnel vision by allowing one to get fresh perspectives from colleagues. Moreover, it generates productivity by allowing different squads to collaborate and share their expertise toward accomplishing a common objective.  

Designers within the UX squad collaborate with fellow designers to resolve a specific design dilemma or challenge. Usually, designers use Slack channels and designated tags to notify the UX and DS/UI squads of a problem. The UX and DS/UI squads sometimes whiteboard and brainstorm the problem together depending on their respective schedules and bandwidth. In most cases, when designers work on a new interface to handle a new workflow, they reach out to as many people on the team as possible to capture feedback to make an informed decision if the designs they created are effective. The design team has a dedicated workflow to submit designs and gather feedback through a work management tool.  
Another crucial aspect of collaboration for the Design Team is to support every individual's professional goals. Along that line of thought, when new employees join the team, they gradually learn about Chatham and its product offerings through the team's tribal knowledge and by collaborating with them on specific projects. In a few situations, the Design Team members also help each other with an area or skillset that they specialize in by providing feedback. For new team members, our team has also curated set of tasks and resources based on their role that can enable them to find their feet quickly entering the new workspace.

We identified a few gaps in our workflow that have impacted the collaboration between and within the squads: 
\begin{enumerate}
\item{\em Conflicting schedules and inconsistent processes for UX meetings hamper collaborations:} Sometimes, the UX and DS/UI squads hesitate to collaborate due to strict deadlines and conflicting schedules. Moreover, sharing background knowledge to provide context to other squad members about the project also takes time and effort. Consequently, fewer collaborations within the UX and DS/UI squads burden the DS/UI squad to ensure consistency within the designed deliverables.  

\item{\em Limited information about the ideation process impacts the effectiveness and time of feedback:} In some projects, the UX squad shares the summarized findings at the end of the ideation process of how they arrived to their designs, which does not help fellow designers to contribute or collaborate earlier in the project. 

\item{\em Challenge of extracting tribal knowledge to identify specific skillset experts in the Design Team:} Besides the set of resources provided to the new team members in the team’s work management platform, team members and new collaborators also felt that it would have helped to ask for suggestions and feedback directly from people who are experts at a certain skill or trait, like UI designing in prototyping software, storytelling, and journey-mapping, without relying on team’s “tribal knowledge”. 
\end{enumerate}
 
To enhance the quality and venues for both inter- and intra-squad collaborations, we are introducing new initiatives. The following initiatives will accommodate each member’s schedule and provide avenues to easily reach specific members of the team who are experts at a particular skill: 

\begin{enumerate}
    \item{\em Dedicated open walk-in office hours to brainstorm in a co-located/hybrid/remote setting:} We decided to dedicate open walk-in office hours for the Design Team, allowing the team members to set objectives for the brainstorming session in advance. This event saves a member's time to look at each team member's calendar to schedule a brainstorming session. 

\item{\em Standardize and templatize the brainstorming process and other UX meetings:} Most of our UX meetings happen virtually. To utilize the dedicated time efficiently, we have started standardizing the UX meetings so that collaborators with diverse backgrounds can effectively collaborate. 

\item{\em Early and frequent sharing of the ideation process of a design solution to gather early and effective feedback:} Our team tries to share the UI interface ideation process with the team as soon as possible. This knowledge is crucial for the UX squad to determine if they have similar design use cases to provide feedback early in the process. Moreover, it helps the squad stay up to date on various design interfaces, components, and use cases that other members are working on.  

\item{\em Create a repository of skills and squad members who are experts and wish to help their fellow team members:} To help new collaborators and other team members reach out to specific experts within the team, we asked squads to create a list of skills for each squad and then asked the squad members to vote on the top three things that they are good at and are willing to be a mentor for.
\end{enumerate}

\section{Conclusion}
\label{sec:conclusion}
In this case study, we have discussed an internal research study carried out by the Design Team at Chatham Financial to better understand the current collaborative practice and ways to foster a conducive environment for collaboration between multifaceted design and product teams to produce impactful end-user experiences. The insights presented are a snapshot of the time while initiatives are continuously being implemented.

\section*{Acknowledgments}
We thank Brett Rodgers, Kris Gabbard, Nevin King, Samantha Andrew, and all our Design and Product colleagues at Chatham for sharing their experiences and input on the case study. 

\bibliographystyle{ACM-Reference-Format}
\bibliography{BIB/references}

\end{document}